\documentclass{aa}  

\usepackage{graphicx}
\usepackage{txfonts}
\begin{document}

   \title{Long-term $K_s$-band photometric monitoring of L dwarfs}

   \subtitle{}

   \author{B. L\'opez Mart\'{\i}
          \inst{1,2}
          \and
          M.~R. Zapatero Osorio\inst{3}
          }

   \institute{
       Centro de Astrobiolog\'{\i}a (INTA-CSIC), P.O. Box 78, E-28261 Villanueva de la Ca\~nada, Madrid, Spain \\              
        \email{belen@cab.inta-csic.es}
         \and
         Saint Louis University -- Madrid Campus, Division of Science, Engineering and Nursing, Avenida~del~Valle 34, E-28003 Madrid, Spain 
         \and
         Centro de Astrobiolog\'{\i}a (INTA-CSIC), Carretera de Ajalvir km. 4, E-28850 Torrej\'on de Ardoz, Spain    
             }

   \date{Received ; accepted }

 
  \abstract
   {
   Ultracool dwarfs ($\lesssim$2500~K) are known to display photometric variability in short timescales    (hours to days), which has usually been related to rotation-modulated dust cloud patterns, or to unresolved companions. 
   }
   {
   We perform photometric time-series analysis of a sample of ten early to mid-L dwarfs in the field over three years of $K_s$-band observations with the OMEGA\,2000 infrared camera of the 3.5m telescope on Calar Alto Observatory between January 2010 and December 2012. This study represents the first systematic long-term photometric monitoring of this kind of object to date.
   }
   {
   We perform $K_s$-band differential photometry of our targets (with typical errors of  $\pm$15-30~mmag  at the 1$\sigma$ level) by subtracting a reference flux from each photometric measurement. This reference flux is computed using three nearby, probably constant stars in the target's field-of-view. We then construct and visually inspect the light curves to search for variability, and use four different periodogram algorithms to look for possible periods in our photometric data. 
  }
   {
   Our targets do not display long-term variability over 1$\sigma$ compared to other nearby stars of similar brightness, nor do the periodograms unveil any possible periodicity for these objects, with two exceptions:  2MASS~J02411151-0326587 and  G196-3B. In the case of 2MASS~J02411151-0326587 (L0), our data suggest  a tentative period of 307$\pm$21~days, at 40\% confidence level, which seems to be associated with peak-to-peak variability of 44$\pm$10~mmag. This object may also display variability in timescales of years, as suggested by the comparison of our Ks-band photometry with 2MASS. For G196-3B (L3), we find peak-to-peak variations of 42$\pm$10~mmag, with  a possible photometric period of 442$\pm$7~days, at 95\% confidence level. This is roughly the double of the astrometric period reported by \citet{zo2014}. Given the significance of these results, further photometric data are required to confirm the long-term variability.
   }
   {   
   Our results suggest that early- to mid-L dwarfs are fairly stable in the $K_s$-band within $\pm$90 mmag at the 3 $\sigma$ level over months to years, which covers hundreds to tens of thousands of rotation cycles. Two out of ten targets show periodic photometric variability at 2.2 $\mu$m with peak-to-peak variations of about 40 mmag and tentative periods of $\sim$300 and $\sim$450 d.
    }


\titlerunning{Long-term $K_s$-band monitoring of L dwarfs}
\authorrunning{B. L\'opez Mart\'{\i} \& M.~R. Zapatero Osorio}

   \maketitle
%
\section{Introduction}\label{sec:intro}

Brown dwarfs have similar physical properties to the coolest and lowest mass stars and to the giant planets; all these groups are termed generically as ``ultracool dwarfs'' ($T_{eff}\lesssim2500$~K). Their most relevant characteristics are their fully convective interiors and the presence of molecules and dust in their atmospheres. For M dwarfs, the most prominent features are the broad TiO and VO absorption bands seen in the red part of their optical spectra. As we proceed towards lower temperature, the refractary elements aggregate into grains and their signature disappears from the spectra of L-dwarfs \citep{tsuji1996, allard1997, kirkpatrick1999, martin1999}. These objects are characterized by a dusty cloud deck in their atmospheres, which also include silicate and iron particles. In addition, their spectra display absorption of metal hydrides (FeH, CrH) and strong alkali lines (K, Na, Rb, Cs), as well as absorption bands of water and CO. For still lower temperatures, these clouds settle below the photosphere and disappear, and methane and water become the most important features in the spectra of T-dwarfs \citep[][and references therein]{kirkpatrick2005}.

Work by many groups over the past decades has shown that ultracool dwarfs display optical and/or infrared photometric variability in timescales of hours or days, usually attributed to rotationally driven cloud instabilities, pulsation, or the presence of unresolved companions \citep[e.g.][and references therein]{bj2005, goldman2005, heinze2013}. However, the hypothesis that some of the reported variability is caused by magnetic activity cannot be ruled out, at least for early-type ultracool dwarfs \citep[e.g.][]{littlefair2008}. On one hand, H$\alpha$ emission, a proxy for chromospheric activity, has been reported even for T-type objects \citep[e.g.][]{burgasser2000}; 
on the other hand, many ultracool dwarfs with spectral types as late as T6.5 are reported to be radio pulsators, a phenomenon usually related to magnetic activity. This radio pulsation has sometimes been related to optical periodic variability \citep[e.g.][]{lane2007,hallinan2008, route2012, harding2013}. Some authors argue, however, that even if some magnetic activity is present, the coupling of the (supposedly weak) magnetic field with the atmosphere will not be very strong, ruling out magnetic cool spots as those detected in solar-type stars \citep[e.g.][]{gelino2002, mohanty2002}. 

Clouds, binarity, or magnetic activity may also yield photometric variability at longer timescales. In solar-type stars, periodic, low-amplitude (a few mmag) photometric variations have been reported, and are generally attributed to spot cycles similar to that of our Sun \citep[e.g.,][]{lockwood1997,radick1998}. In ultracool dwarfs, however, owing to their dusty atmospheres,  it seems more likely that long-term variations are produced by weather phenomena, probably resembling those observed in the giant planets of our own Solar System. In the case of planets, long-term periodic photometric variations are expected to be associated with global changes in the planet's belts and zones, and with seasonal cycles modifying the large-scale cloud structures \citep[e.g.][]{slv1991, slv2011, slv1996}. Although little work has been done so far in quantifying these effects, these variations are expected to have low amplitudes (not more than a few percent; R. Hueso, private communication) and cover timescales of one to two years in longer cycles of ten to 30 years; such variability has been reported at least for Uranus and Neptune, with optical ($B$-band)  peak-to-peak variations of about 25 and 7~mmag, respectively, over several decades \citep[][]{lockwood2006}. Irregular and brusque variations caused by storms or structural changes in the planet's belts have also been observed in giant planets \citep[e.g.][]{tejfel1994, slv1994, rogers2009, fletcher2011, ph2012}, and they are likely to happen in ultracool dwarfs as well. 

Photometric data from different epochs, separated by months, are already available for a few ultracool dwarfs \citep[e.g.][]{metchev2013}, suggesting that these objects may indeed display variability at these timescales. However, to our knowledge no systematic long-term monitoring of these objects had been carried out so far. This paper presents a study of long-term (from three months to three years) variability of ten nearby ultracool dwarfs. Our work is thus the first prospect exploring variability in this time range at such low temperatures.

The structure of the paper is as follows: Section~\ref{sec:targets} presents a brief description of our targets. The procedures to obtain differential and absolute photometry from our images are explained in Section~\ref{sec:phot}. Section~\ref{sec:period} describes the periodogram analysis carried out on our data. We discuss our results in Section~\ref{sec:res}. Finally, Section~\ref{sec:concl} summarizes our conclusions.

\section{Target description}\label{sec:targets}

\begin{table*}[t]
\label{tab:targets}
\caption{Target properties$^{\mathrm{a}}$}
\centering
\begin{tabular}{l c c c c c c c}
\hline
\hline\noalign{\smallskip}
  \multicolumn{1}{l}{Object} &
  \multicolumn{1}{c}{SpT} &
  \multicolumn{1}{c}{$d$} &
  \multicolumn{1}{c}{Age range} &
  \multicolumn{1}{c}{Mass} &
  \multicolumn{1}{c}{$\bar{K_s}^{\mathrm{b}}$} &
  \multicolumn{1}{c}{$\sigma_{K}\,^{\mathrm{c}}$} &
\multicolumn{1}{c}{2MASS $K_s$} \\
  \multicolumn{1}{l}{} &
  \multicolumn{1}{c}{} &
  \multicolumn{1}{c}{(pc)} &
  \multicolumn{1}{c}{(Myr)} &
  \multicolumn{1}{c}{($M_J$)} &
  \multicolumn{1}{c}{(mag)} &
  \multicolumn{1}{c}{(mag)} &
  \multicolumn{1}{c}{(mag)} \\
  \noalign{\smallskip}\hline\noalign{\smallskip}
  \object{2MASS~J00332386$-$1521309}  & L4$\beta$        & 40$\pm$4 &  $\le$10    & 4 &  13.439$\pm$0.011     & $\pm$0.04 & 13.410$\pm$0.039 \\
  \object{2MASS~J00452143+1634446} & L2$\beta$        & 17.5$\pm$0.6 &  10-100    & 15 &  11.326 $\pm$0.005   & $\pm$0.02 & 11.366$\pm$0.021 \\
  \object{2MASS~J02411151$-$0326587}  & L0$\gamma$  & 47$\pm$6 & $\ge$500    & 80 &  14.175$\pm$0.005 & $\pm$0.018 & 14.04$\pm$0.05 \\
  \object{2MASS~J03552337+1133437} & L5$\gamma$  &   9.0$\pm$0.4 &   50-500   & 23 &  11.531$\pm$0.029      & $\pm$0.12 & 11.526$\pm$0.021 \\
  \object{2MASS~J05012406$-$0010452}  & L4$\gamma$  & 19.6$\pm$1.4 & 50-500   & 25 &  12.952$\pm$0.012     & $\pm$0.05 & 12.963$\pm$0.035 \\
  \object{G196$-$3B}                                        & L3$\beta$       & 24$\pm$2 &  10-300    & 15 &  12.795$\pm$0.004    & $\pm$0.02 & 12.778$\pm$0.034 \\
  \object{2MASS~J10224821+5825453} & L1$\beta$          & 21.6$\pm$0.6 & 100-1000 & 50 & 12.112$\pm$0.008 & $\pm$0.04 & 12.160$\pm$0.025 \\
  \object{2MASS~J15525906+2948485} & L0$\beta$       & 21.0$\pm$0.4 & $\ge$500     & 75 &  12.028$\pm$0.007    & $\pm$0.04 & 12.022$\pm$0.028 \\
  \object{2MASS~J17260007+1538190} & L3$\beta$       & 35$\pm$4 &  10-300     & 20 &  13.748$\pm$0.004     & $\pm$0.02 & 13.659$\pm$0.050\\
  \object{2MASS~J22081363+2921215} & L3$\gamma$ & 47.2$\pm$1.6 &  10-300      & 15 & 14.092$\pm$0.005     & $\pm$0.03 & 14.148$\pm$0.073 \\
  \noalign{\smallskip}\hline\noalign{\smallskip}
\end{tabular}
\begin{flushleft}
{\bf Notes.} \\
     $^{\mathrm{a}}$ References: \citet{cruz2009, zo2014}; this work. \\
     $^{\mathrm{b}}$ Mean $K_s$-band magnitude measured in this work. The error is the standard error of the mean, calculated using Eq.~\ref{eq:errphot} \\
     $^{\mathrm{c}}$ Standard deviation of the $K_s$-band measurements from this work.
 \end{flushleft}
\end{table*}

The targets studied in this work are the same objects presented in the parallax study by \citet{zo2014}: ten ultracool field dwarfs of spectral types L0 to L5 and spectroscopic evidence for low-gravity atmospheres \citep[][and references therein]{cruz2009,zo2014}. The trigonometric parallaxes measured by \citet{zo2014} confirmed that they are located at nearby distances (9-47~pc). In the same work, seven of the objects were confirmed to be young (10-500~Myr) and substellar (4-25~$M_J$) by their location in the Hertzprung-Russell diagram, while three of them ( \object{J0241-0326},   \object{J1022+5825} and  \object{J1552+2948}) were shown to be probably older than 500~Myr and to have masses closer to or above the lithium burning mass limit. Table~1 
summarizes the relevant information about our objects.

The analysis of the multi-epoch astrometric data ($d\alpha$) revealed no companions with masses above 25$M_J$ in face-on, circular orbits with periods between 60-90 days and three years around any of the targets. The only possible exception was G196-3B, for which a tentative signal corresponding to a period of about 228.46 days was detected, hinting at a similar-mass double system with a separation of 0.27 AU \citep[for details, see][]{zo2014}. However, the possible binary nature of G196-3B needs to be confirmed with additional observations.

\section{Data processing}\label{sec:phot}

   \begin{figure}[t]
   \centering
     \includegraphics[width=10.5cm]{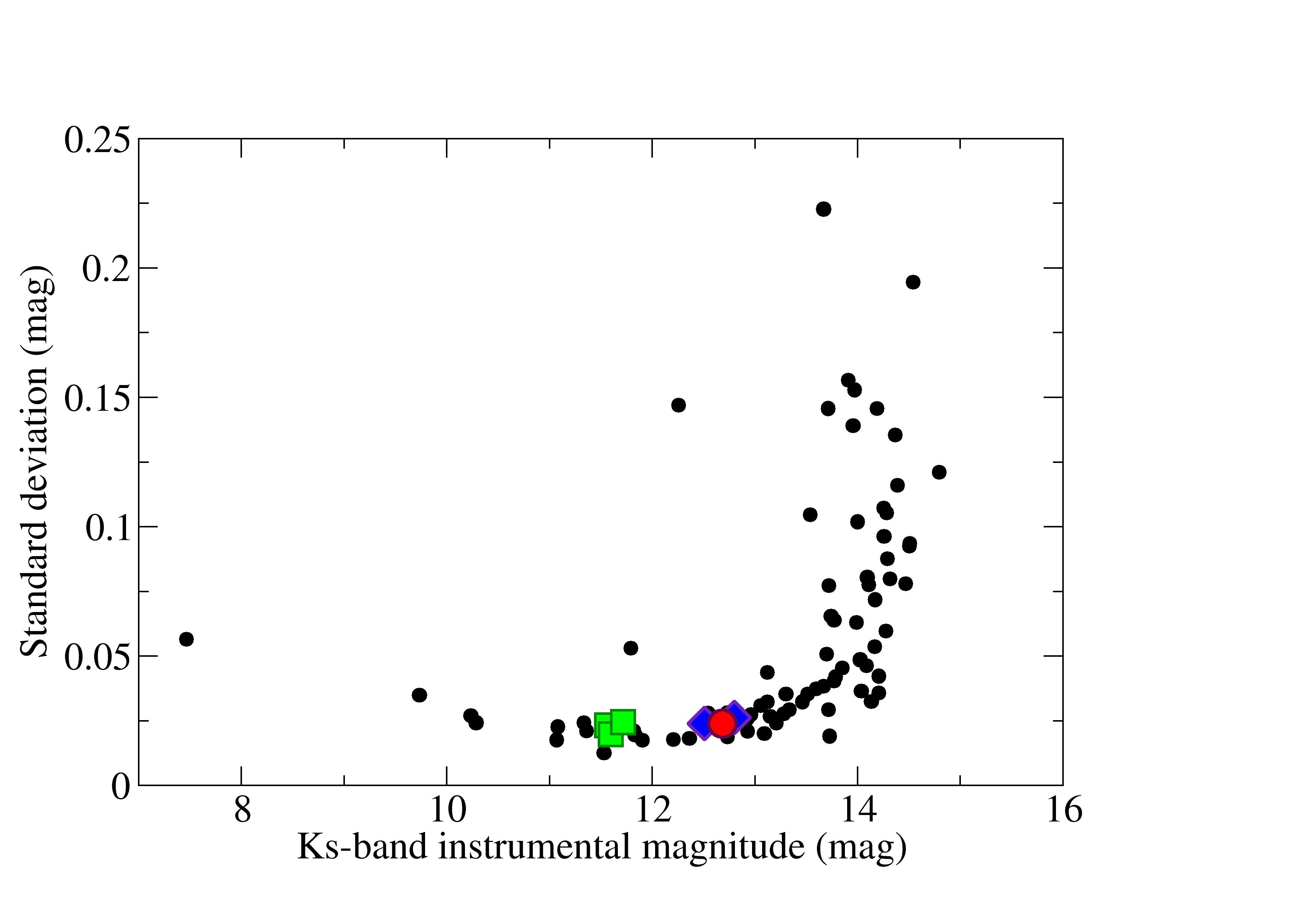}\hfill
     \caption{\footnotesize
	       Example of the kind of diagram used to select the reference and comparison stars. The plot shows the standard deviation of the differential magnitudes of the stars in the \object{J0241-0326} field, computed with respect to a large sample of these stars, versus the instrumental magnitude (see text for details). Variable stars stand out in this diagram because of their large standard deviations compared to other stars of similar brightness. Our target is indicated with a red circle, and the finally selected reference and comparison stars are plotted with green squares and blue diamonds, respectively.
	       	       }
         \label{fig:sigma}
   \end{figure}

\subsection{Instrumental photometry}\label{sec:insphot}

For our photometric time-series analysis, we made use of the same multi-epoch Ks-band images used by \cite{zo2014} in their astrometric study. We refer to that work for a complete description of the data, including the observing log, and of the data reduction. We only considered the images taken with the wide-field (15'$\times$15') OMEGA\,2000 camera installed in the prime focus of the 3.5m telescope of Calar Alto Observatory (Almer\'{\i}a, Spain). Our observations were carried out in the period between 2010 January to 2012 December, typically once per month; thus, our time baseline is 3~yr. 
However, owing to a technical failure of the telescope, the observations were interrupted for eight months between 2010 August and 2011 March. The number of images per target ranged between 12 and 27. The data were taken and processed in a standard way, as explained in  \cite{zo2014}.

Automatic object search and aperture photometry on the OMEGA2000 Ks-band images were performed using SExtractor \citep{bertin1996} in a two-step procedure: First, we ran SExtractor with default parameters to identify the brightest objects in each image, and to measure their full-width at half-maximum (FWHM). The average FWHM for each image was then used as the aperture radius to perform the photometry of all the objects in the image in the second SExtractor run; values ranged between 1.7 and 4$\arcsec$, depending on night conditions. Both the detection and analysis threshold were set to 5$\sigma$ over background. In this second run, fluxes (in counts) as well as magnitudes were stored.

\subsection{Differential photometry}\label{sec:difphot}

\begin{table*}[t]
\label{tab:stdev}
\caption{
Magnitude differences $\Delta K_S$  (defined as $K_{star}-K_{target}$) with respect to the targets and standard deviations $\sigma(dK)$ of the light curves for our targets and their comparison and reference stars.}
\centering
\begin{tabular}{l c r c r c r c r c r c}
\hline
\hline\noalign{\smallskip}
\multicolumn{1}{l}{Field} &    
\multicolumn{1}{c}{Target} &
\multicolumn{2}{c}{Comparison 1} & 
\multicolumn{2}{c}{Comparison 2} &
\multicolumn{2}{c}{Reference 1} &
\multicolumn{2}{c}{Reference 2} &
\multicolumn{2}{c}{Reference 3} \\
\noalign{\smallskip}
\multicolumn{1}{l}{} &    
\multicolumn{1}{c}{$\sigma(dK)$} & 
\multicolumn{1}{c}{$\Delta K_S$} &
\multicolumn{1}{c}{$\sigma(dK)$} &
\multicolumn{1}{c}{$\Delta K_S$} &
\multicolumn{1}{l}{$\sigma(dK)$} &    
\multicolumn{1}{c}{$\Delta K_S$} &
\multicolumn{1}{c}{$\sigma(dK)$} & 
\multicolumn{1}{c}{$\Delta K_S$} &
\multicolumn{1}{c}{$\sigma(dK)$} &
\multicolumn{1}{c}{$\Delta K_S$} &
\multicolumn{1}{c}{$\sigma(dK)$} \\
\multicolumn{1}{l}{} &    
\multicolumn{1}{c}{(mmag)} & 
\multicolumn{1}{c}{(mag)} &
\multicolumn{1}{c}{(mmag)} &
\multicolumn{1}{c}{(mag)} &
\multicolumn{1}{l}{(mmag)} &    
\multicolumn{1}{c}{(mag)} &
\multicolumn{1}{c}{(mmag)} & 
\multicolumn{1}{c}{(mag)} &
\multicolumn{1}{c}{(mmag)} &
\multicolumn{1}{c}{(mag)} &
\multicolumn{1}{c}{(mmag)} \\
\noalign{\smallskip}\hline\noalign{\smallskip}
\object{J0033$-$1521}  & $\pm$22 &    0.097 & $\pm$19  & $-$0.005 & $\pm$24  & $-$0.720 & $\pm$21  & $-$0.647 & $\pm$13  & $-$0.798 & $\pm$20  \\
\object{J0045+1634}    & $\pm$15 &    0.024 & $\pm$44  &    0.596 & $\pm$36  & $-$0.961 & $\pm$22  & $-$1.079 & $\pm$26  & $-$0.551 & $\pm$26  \\
\object{J0241$-$0326}  & $\pm$19 &    0.290 & $\pm$13  &    0.352 & $\pm$23  &    0.841 & $\pm$16  &    0.826 & $\pm$11  &    0.736 & $\pm$14 \\ 
\object{J0355+1133}    & $\pm$18 & $-$1.369 & $\pm$11  &    0.152 & $\pm$16  & $-$0.984 & $\pm$22  & $-$0.920 & $\pm$23  & $-$0.848 & $\pm$15 \\ 
\object{J0501$-$0010}  & $\pm$31 & $-$0.003 & $\pm$29  &    0.062 & $\pm$24  &    0.150 & $\pm$21  & $-$0.173 & $\pm$18  &    0.246 & $\pm$22  \\
\object{G196$-$3B}     & $\pm$23 & $-$0.112 & $\pm$17  &    0.232 & $\pm$19  & $-$0.320 & $\pm$21  & $-$0.315 & $\pm$20  &    0.196 & $\pm$33  \\ 
\object{J1022+5825}    & $\pm$21 &    0.607 & $\pm$11  & $-$0.352 & $\pm$27  &    0.712 & $\pm$21  &    0.770 & $\pm$22  &    0.801 & $\pm$23  \\  
\object{J1552+2948}    & $\pm$18 & $-$0.004 & $\pm$20  &    0.007 & $\pm$22  &    0.055 & $\pm$30  &    0.414 & $\pm$20  & $-$0.533 & $\pm$27  \\
\object{J1726+1538}    & $\pm$16 &    0.045 & $\pm$16  &    0.052 & $\pm$19  & $-$0.486 & $\pm$17  & $-$0.467 & $\pm$11  & $-$0.406 & $\pm$14 \\ 
\object{J2208+2921}    & $\pm$40 & $-$0.066 & $\pm$32  &    0.031 & $\pm$43  & $-$3.731 & $\pm$29  & $-$4.070 & $\pm$19  & $-$3.703 & $\pm$15 \\
\noalign{\smallskip}\hline\noalign{\smallskip}
\end{tabular}
\end{table*}

Differential photometry was performed with respect to three reference stars in each field, which were used to compute the reference flux to be subtracted from the target flux, as explained below.  These stars were preferably chosen to be brighter than the target itself, but this was often difficult to accomplish, because our targets are among the brightest objects seen in the OMEGA\,2000 field-of-view. Therefore, we sometimes had to take stars of similar brightness ($\pm0.5$~mag) or even slightly fainter (up to 1~mag) than our target (e.g. for \object{J0045+1634}). In addition, for each target,  two comparison stars of similar brightness were selected; these stars were used to check that any apparent variability was not caused by instrumental effects. Most comparison and reference stars are seen within 5$\arcmin$ from the corresponding target, in any case never further than 7$\arcmin$.

A very important requirement for both reference and comparison stars is that they are non-variable. To ensure that we selected appropriate objects, we proceeded as follows: First, a differential magnitude was computed for every detected star and observation with respect to a large number of stars in the same field. The only condition imposed on these stars was that they should be relatively bright (in order to have enough signal-to-noise to ensure a good flux measurement) but still lie in the linearity range of the detector. For each star and observation, we used the relation:

\begin{equation}\label{eq:difphot}
dK=-2.5\cdot\log (F-\frac{1}{N}\sum_{i=1}^{N}{F_i}),
\end{equation}

\noindent
where $F$ is the flux of the star, $N$ is the total number of stars considered in the field,  and $F_i$ is the flux of the $i$-th star in this sample of $N$ stars. The average of all $F_i$ fluxes provides a reference flux that is subtracted from the star's flux, and the resulting differential flux is converted into a magnitude. The number $N$ of stars used greatly varied depending on the field, from seven
for \object{J0045+1634} to 78 for \object{J0241-0326}.

We then computed the standard deviations of the so-obtained differential magnitudes over the whole observing period for all the stars detected in each field, and plotted them versus the instrumental magnitudes measured in the same observation. As discussed in \citet{caballero2004}, this diagram allows us to easily identify possible variable stars, which stand out immediately because of their high standard deviation values as compared with other stars of similar brightness.  As an example, Fig.~\ref{fig:sigma} shows the diagram for \object{J0241-0326}. By inspecting these diagrams, we ensured that the selected reference and comparison stars were likely to be non-variable, compared to the rest of stars of similar brightness in the same field. In the linearity range of the detector, typical standard deviations for constant stars had values $\lesssim$5~mmag. However, a few stars stood out with standard deviations as high as two or three times the typical values for stars of similar brightness.

The reference stars selected in this way were used to compute accurate differential magnitudes for the targets and their comparison stars using Eq.~\ref{eq:difphot}. Note that the sum now included only the fluxes of the three selected, well-behaved reference stars ($N=3$). Table~2 
lists the magnitude differences $\Delta K_S$ between the targets and their respective comparison and reference stars ($\Delta K_S=K_{star}-K_{target}$), and the $\sigma(dK)$ values for all of them. 
In the rest of the paper, we will refer to the differential photometry computed in this way. We will not use the differential photometry computed using a large sample of stars, because it included objects that are likely to be variable. The purpose of the above exercise was only to discard obvious variable stars from the reference and comparison samples. The light curves of our targets are discussed in Section~\ref{sec:curveres}.

As a sanity check, we also computed the differential fluxes and magnitudes using only two from the three reference stars. The resulting light curves for all possible combinations were visually inspected to confirm that the use of one particular star was not introducing some artificial variability in the data. The comparison concluded that all light curves for the same object (target or comparison star) showed very similar shapes, regardless of the stars used as reference to construct them. We also inspected the light curves lof the reference stars themselves (constructed using the other two stars as reference), and found no evident variability in them: their $\sigma(dK)$ values are similar to what is found for the light curves of the comparison stars. 

   \begin{figure*}[t]
   \centering
     \includegraphics[width=1.1\textwidth]{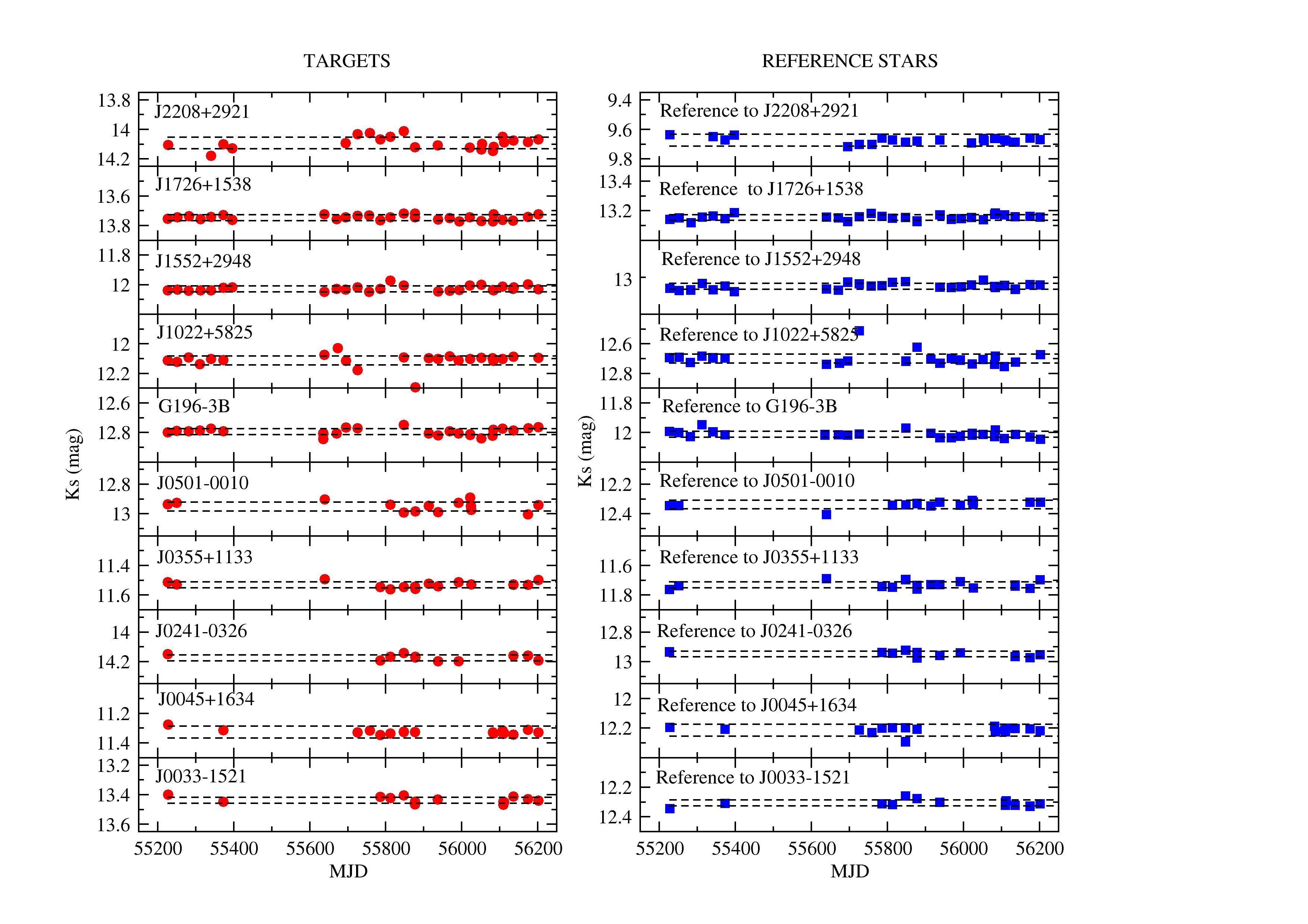}\hfill
     \caption{\footnotesize
	       Photometric light curves for our ten targets (left panels), and for one of the reference stars used to compute the differential photometry for each target (right panels). The differential photometry (red circles and blue squares, respectively) has been shifted so that the average value corresponds to the mean Ks-magnitude of the objects. The horizontal dashed lines show the $\pm1\sigma$ variation with respect to the average values for two comparison stars in the field of view of each target and reference star.
	       }
         \label{fig:curves}
   \end{figure*}

\subsection{Absolute photometry}\label{sec:absphot}

The instrumental photometry was calibrated with respect to the 2~Micron All-Sky Survey (2MASS) system \citep{skrutskie2006} using all the 2MASS sources present in the field to derive a zero point for each image; this zero point was then subtracted from the instrumental magnitudes. The so-calibrated magnitudes were averaged to derive a mean $K_s$-band magnitude for each target, as listed in Table~1. 
In each case, the standard error of the mean, $\sigma_{\bar{K}}$, was computed from the standard deviation of the absolute magnitudes $\sigma_{K}$ (typically between 0.02 and 0.04~mag; see Table~1) 
and the number of observations $N_{obs}$ as

\begin{equation}\label{eq:errphot}
\sigma_{\bar{K}}=\frac{\sigma_{K}}{\sqrt{N_{obs}}}
\end{equation}

For comparison, Table~1 
also lists the published 2MASS $K_s$-band magnitudes for our objects. In general, our mean magnitudes agree well with the 2MASS values within the quoted errors and standard deviations. The magnitude differences between 2MASS and our measurements are also in good agreement with the $\sigma(dK)$ values for a constant star estimated from the differential photometry of the comparison stars in every target field, meaning that the differences are not significant. 
However, the differences between our average magnitudes and the 2MASS values are certainly larger than the estimated errors and $\sigma(dK)$ values for at least one object, namely \object{J0241-0326} (0.135~mag, more than 2$\sigma$ difference). This suggests that this object may be variable in timescales larger than considered in the present study, as the time span between the 2MASS observation and our OMEGA\,2000 photometry is 12 to 14 years. Interestingly, \object{J0241-0326} is also one of our few targets displaying indications of possible variability (see Sect.~\ref{sec:res}).

\section{Periodogram analysis}\label{sec:period}

To test the periodicity of our targets, we made use of the online periodogram analysis tool from the NASA Exoplanet Archive\footnote{\footnotesize http://exoplanetarchive.ipac.caltech.edu/applications/Periodogram/}. The service provides periodicity analysis using three different algorithms, each suitable for a particular type of variability. The Lomb-Scargle periodogram \citep[LS;][]{scargle1982} is an approximation of the Fourier transform for unevenly spaced time sampling; therefore, it is best suited to identify sinusoidal periodic signals.  The Box Fitting Least Squares algorithm \citep[BLS;][]{kovacs2002} fits the time-series to a repeating box-shaped curve, and is thus optimized for the detection of transit events. We note, though, that such events are unlikely to be detectable with the time sampling of our observations, because a transit event in a timescale of months would only be possible with highly wide-separation binary systems, which are extremely rare among ultracool dwarfs \citep{allen2007b}. The third available algorithm is the Plavchan periodogram \citep{plavchan2008}, a binless implementation of the phase dispersion minimization approach of \citet{stellingwerf1978}. It identifies periods with coherent (i.e. smoothest) phased time-series curves, with no assumption about the shape of the underlying periodic signal. Hence, if a peak appeared in either the LS or the BLS periodogram, we expected to confirm it with the Plavchan algorithm.

The range of periods tested by the algorithms spanned from about 60 to about 1064 days (three to 35 months). We ran these algorithms not only on the light curves of our targets, but also on the light curves of the comparison and reference stars in exactly the same way as for the targets. This provided further evidence that none of these objects displayed significant variability (within our quoted errors) at the timescales we are probing, and are thus suitable for the purposes we are using them.

For those objects showing periodogram peaks, we used the CLEAN algorithm \citep{roberts1987} implemented in the Starlink PERIOD time-series analysis package to further test the reliability of those peaks. This algorithm is especially useful for unequally spaced data, providing a simple way to understand and remove the artifacts introduced by missing data. The algorithm was run with five iterations and a loop gain of 0.2 on a previous LS periodogram derived with the same package.  The LS periodograms obtained with PERIOD are consistent with those of the NASA Exoplanet Archive tool for the same algorithm. The CLEANed periodograms, on the other hand, provide slightly different periods compared to the simple LS periodograms, and some peaks are not present as a result of the removal of artifacts. Neither the Plavchan nor the BLS algorithm are implemented in the PERIOD package.

The results from the periodogram analysis are discussed in Section~\ref{sec:periodres}.

   \begin{figure*}[t]
   \centering
     \includegraphics[width=1.2\textwidth]{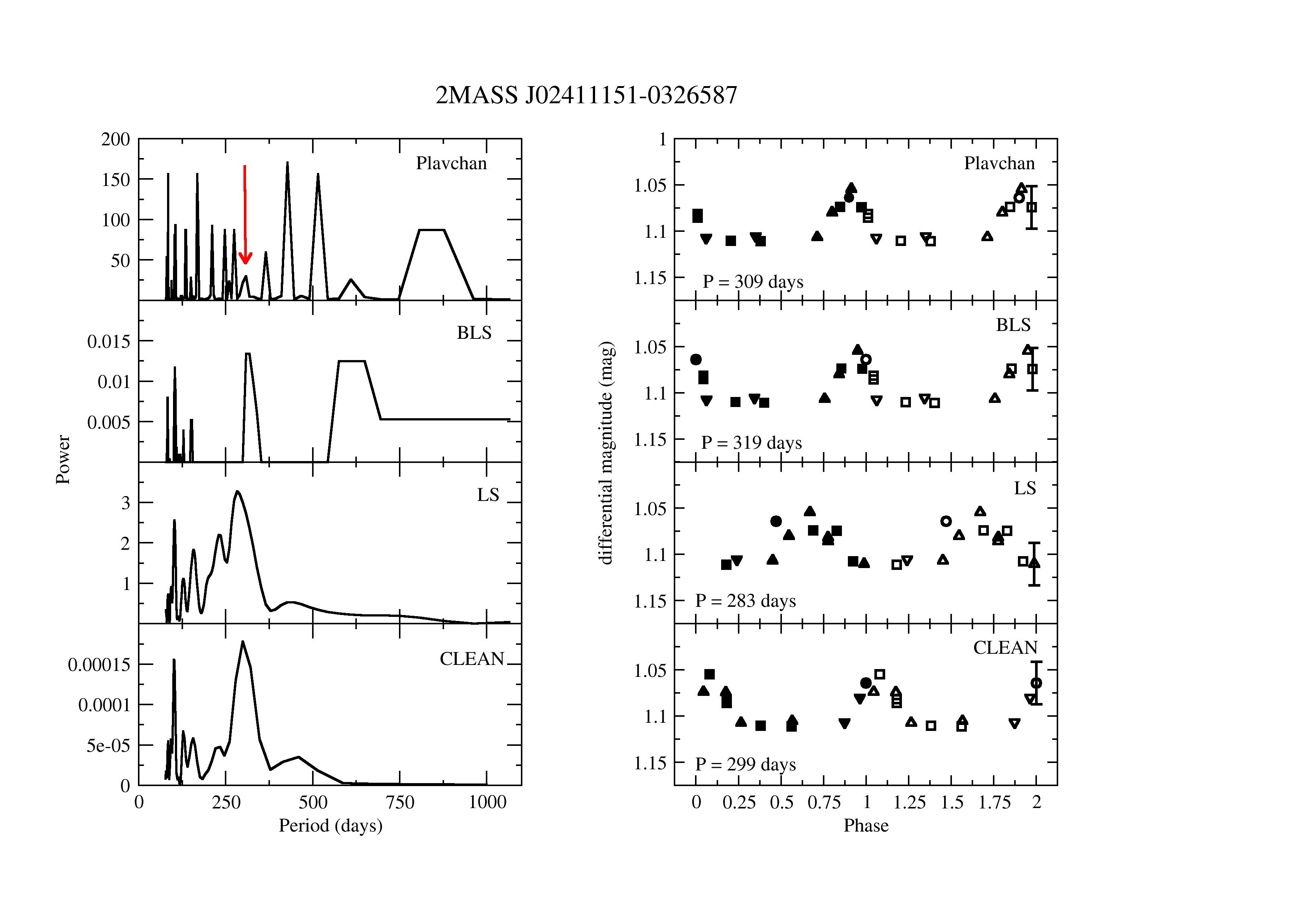}\hfill
     \caption{\footnotesize
	       Periodograms (left panels) and phased curves corresponding to their most significant peaks (right panels) for \object{J0241-0326}. In the phased curves, different symbol shapes indicate data points corresponding to the different cycles, while filled and open symbols of the same shape correspond to duplicated data points. The typical error bar is shown for the last data point. The red arrow in the upper left panel indicates the peak in best agreement with the highest peak in the rest of periodograms (see text for discussion). 
	       	       }
         \label{fig:j0241}
   \end{figure*}

   \begin{figure*}[t]
   \centering
     \includegraphics[width=1.2\textwidth]{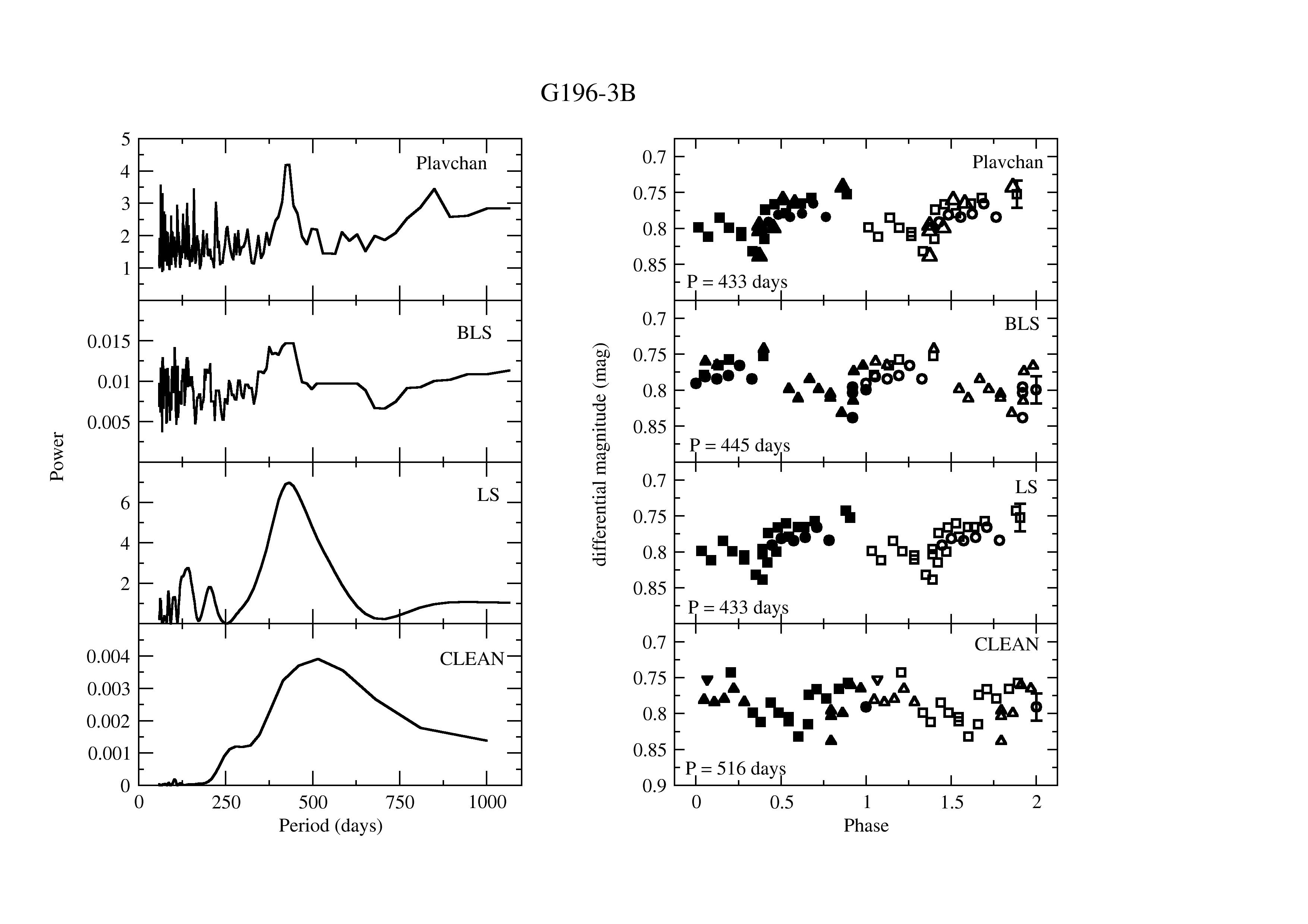}\hfill
     \caption{\footnotesize
Periodograms (left panels) and phased curves corresponding to their most significant peaks (right panels) for \object{G196-3B}. Symbols as in Fig.~\ref{fig:j0241}. The typical error bar is shown for the last data point.
		       }
         \label{fig:g196-3b}
   \end{figure*}

\section{Results and discussion}\label{sec:res}

\subsection{Irregular variability} \label{sec:curveres}

The light curves of our targets are shown in the left panels of Fig.~\ref{fig:curves}. For better understanding, the differential magnitudes are shifted, so that the mean value corresponds to the mean $K_s$-band magnitude listed in Table~1, 
computed as explained in Sect.~\ref{sec:absphot}. The horizontal dashed lines indicate the standard deviation of the differential magnitudes, $\sigma(dK)$, for a ``constant'' star of similar brightness as our targets, which is estimated from the light curves of the corresponding comparison stars. These $\sigma(dK)$ values, which represent the expected photometric error for our targets, are typically $\pm$15-30~mmag (see Table~2). 

None of our targets display any photometric measurement deviating by more than 3$\sigma$ from their mean $K_s$ magnitudes. Actually, as shown in Table~2, 
the standard deviations of the differential photometry measurements for our targets $\sigma(dK)$ coincide with those of the comparison stars in their fields of view. This suggests that no irregular variability of high photometric amplitude is seen in any of our targets during the time interval of the observations. The right panels of Fig.~\ref{fig:curves} show the light curves of one reference star per target, proving that the reference stars remain stable during the time interval of our study. 
We thus conclude that, at a confidence level of 99\%, none of our early L-dwarfs show photometric variations with amplitudes higher than 90 mmag.

\subsection{Periodic variability} \label{sec:periodres}

Even at $1\sigma$, the periodogram analysis of the light curves may pinpoint the existence of periodic variability \citep[e.g.][]{bartlett2009}. However, for most of our targets, the algorithms unveiled no significant peaks, leading us to conclude that the majority of our objects are not periodically variable at the timescales that we can explore.  A peak was considered significant if it had a p-value $<$ 0.1, where the p-value is defined as the probability that random noise will produce a peak with the same power value as the observed peak. With this criterion,  we found a peak of possible significance in one or more of the periodograms in only four cases, namely  \object{J0241-0326},  \object{G196-3B}, \object{J1552+2948}, and  \object{J2208+2921}. For two of these objects, \object{J0241-0326} and \object{G196-3B}, there seems to be a rough agreement in at least three of the four periodograms (see discussion below).  For the other two objects (\object{J1552+2948} and  \object{J2208+2921}), however, there was no overall agreement among the peaks found by the different periodograms, and the folded light curves according to the periods suggested by the algorithms did not display any clear variability pattern. We thus conclude that the detected peaks for these two objects are probably artifacts caused by the time span and irregular sampling of our observations, and that no variability can be deduced for them from our data.

The periodograms for the two objects displaying possible periodic variability are shown in the left panels of Figs.~\ref{fig:j0241} and \ref{fig:g196-3b}. The right panels show the phased light curves corresponding to the most significant period found with each algorithm. We will now discuss these objects in detail.

\subsubsection{\object{2MASS~J02411151-0326587} (\object{J0241-0326})}\label{sec:j0241}

The object \object{J0241-0326}  displays a clear peak at 319$\pm$33~days in the BLS periodogram  (p-value$=$0.048, computed for a sample of 120 periods and a Gaussian power distribution; second upper panel on the left in Fig.~\ref{fig:j0241}), where the error has been calculated as the FWHM of the peak. Also the LS periodogram (second lower panel on the left) displays a similar, though broad, peak, which would correspond to a possible period of 283$\pm$55~days in this case. A similar value (299$\pm$49~days) is found in the CLEANed periodogram 
(lowest left panel), 
fully consistent (at the 1$\sigma$ level) with the peaks found by the BLS and LS algorithms.  The CLEAN and LS periodograms actually look very similar in shape. 

The Plavchan periodogram (highest left panel of Fig.~\ref{fig:j0241}), on the other hand, displays a number of high peaks between 140 and 500 days that have no equivalent with the rest of the algorithms. This suggest that such peaks do not correspond to real periods, but are most probably artifacts. Interestingly, if such peaks are not considered, the most relevant feature beyond 200 days turns out to be a lower peak at 309$\pm$8~days, fairly coincident with the peaks seen in the other periodograms. This peak is indicated with a red arrow in Fig.~\ref{fig:j0241}. 

We computed the significance level of the resulting LS peak against the hypothesis of random noise and found it to be 40\%. This value is low because of the many relatively high peaks at shortest periods, and because of the low number of data points. Even so, the fact that a prominent peak is found around very similar values in at least three of the four tested periodograms gives some credibility to the existence of periodicity in the data. Furthermore, no similar peaks are found in any of the comparison stars, which have $p$-values $\ge$\,0.2 for their most significant peaks (computed from the BLS algorithm as for the target), thus supporting the $\sim$300-days period of J0241$-$0326.

The phased curves corresponding to the possible period values associated with the highest peak from each periodogram are shown in the right panels of  Fig.~\ref{fig:j0241}, except in the case of the Plavchan algorithm, where we show the curve folded according to the 309 days period. These curves indeed suggest periodic peak-to-peak luminosity variation of 44$\pm$10~mmag. However, their exact shapes vary slightly depending on the period value used to fold the light curves (a result not at all unexpected), with the LS  algorithm yielding a fairly sinusoidal curve and the rest of algorithms suggesting a longer quiescent phase between relatively fast increases and drops of luminosity. Clearly, our data are not sufficient to constrain the shape of the light curve. Therefore, we refrain from providing any physical interpretation of the possible variability of this object until an improved dataset is available. 

On the other hand, we recall that \object{J0241-0326} is the only object in our sample showing
relatively large disagreement between the 2MASS $K_s$ magnitude and our OMEGA\,2000 $K_s$-band photometry: As shown in Table~1, 
the 2MASS measurement, which was taken in 1998, is about 0.1~mag brighter than our mean $K_s$ magnitude, suggesting variability at even longer timescales ($\sim$10~yr) than the ones probed in this work. Follow-up observations may confirm this behaviour in the very long term.

We also note that \citet{zo2011} reported linear polarization in the J-band for \object{J0241-0326}, which they related to the possible presence of dust, as this object is overluminous in the infrared ($\gtrsim$2~$\mu$m) compared to other L0/L1 dwarfs. Follow-up observations of this object suggest that the linear polarization in the J-band is variable \citep{miles2014}. If confirmed, it would be interesting to study the possible link between the photometric variability and the variations in linear polarization for \object{J0241-0326}.

\subsubsection{\object{G196-3B}}\label{sec:g196-3b}

The object \object{G196-3B} presents a clear peak at 433~days in both the Plavchan periodogram (433$\pm$19~days; highest left panel of Fig.~\ref{fig:g196-3b}) and the LS periodogram (433$\pm$100~days; second lower panel on the left). In the latter case, the peak is quite broad and slightly assymmetric, being a bit broader towards longer times. A similar peak is also observed in the CLEANed periodogram (516$\pm$276~days; lowest left panel); however, the found peak in this case is remarkably broader (thus less reliable) than in the LS and Plavchan periodograms. The BLS periodogram (second upper left panel) does not display clear peaks, although a much less significant maximum is found around a similar value, 445$\pm$12~days. This may be caused by the fact that the BLS algorithm is not well suited to detect sinusoidal variations. Contrary to J0241$-$0326, all LS, Plavchan, and BLS periodograms of G\,196$-$3B provide the lowest p-values (computed for a total of 270 periods and different power distributions) with the peak at 433-445~days. Particularly, we determine the statistical significance level of the LS peak at 433~days to be 95\%.

A period peak value of 445~days is also found with the LS algorithm comparison stars (with errors of around 70~days). However, the 445~days period of the comparison stars is not confirmed by any of the other three algorithms. Moreover, the p-values obtained for the 445~days peak in the LS periodograms of these stars, which were also computed for a total of 270 periods and the same power distributions as for the target, are two orders of magnitude higher than that of G\,196$-$B ($p$ = 0.0017) and have less statistical significance ($<$50\%). No peak around 430-450~days is seen in the LS periodograms of two of the three reference stars.

For  \object{G196-3B}, all the phased light curves shown in the right panels of Fig.~\ref{fig:g196-3b}, which correspond to the most significant peak in each periodogram, look fairly sinusoidal, with peak-to-peak variations of 42$\pm$10~mmag. The periodic variation is least evident in the CLEANed phase curve (lowest right panel of Fig.~\ref{fig:g196-3b}) compared to the other three, a reflection of the much more uncertain period value. However, without further observations, the detection of variability in G196-3B remains tentative. 

Interestingly, the astrometric analysis of these same data suggested a periodic variation of 228.46 days for  \object{G196-3B}, a value nearly one-half of the maximum peak value we find with the periodogram analysis \citep{zo2014}. This could indicate that both values are harmonics of the same physical period, which may be related to an unsolved companion. However, we cannot rule out the possibility that the astrometric period is actually induced by the photometric variation (if confirmed).

\section{Conclusions and final remarks}\label{sec:concl}

In this paper, we investigated the long-term $K_S$-band photometric variability of a sample of ten field L-dwarfs. Our data covered a range of three years with a typical rate of one observation per month. We performed aperture photometry and computed differential photometry for our targets using three bright reference stars near each object, with typical photometric errors of $\pm$15-30~mmag. We then performed a periodogram analysis using four different algorithms (Plavchan, BLS, LS, and CLEAN) to search or periodicity in our photometric data.

The rotation periodicity of our targets is not available in the literature. However, L dwarfs are known to be fast rotators with spectroscopic rotational velocities ranging from $\sim$10 through $\sim$90 km\,s$^{-1}$ at nearly one-third of the break-up velocity \citep[][and references therein]{konopacky2012}. By adopting a radius of 0.1--0.3 R$_\odot$, which is the size predicted for L-type sources with ages between 50 Myr and a few Gyr by evolutionary models (Chabrier et al. 2000), the expected rotation periods of our targets lie in the interval 1--35 h. Our data do not sample these short periods of time, rather they cover hundreds ($>$750) to tens of thousands of rotation loops. Part of the photometric $K_s$ dispersion seen in the derived light curves may be due to observing the targets at different phases of their rotation loops. From the notorious stability of our data, we conclude that any long-term cycle or irregular variability with amplitudes $>$90 mmag and timescales of 60--90 days through three years can be discarded with a confidence level of 99\%. This implies that any weather phenomena and/or magnetic activity present in these dwarfs do not modify the objects' brightness at 2.2 $\mu$m significantly over thousands of rotation cycles. 

At a lower confidence level and smaller light curve amplitude, the periodogram analysis suggested that at least two objects in our sample display long-term periodic variability that might be associated with weather or magnetic cycles: In the case of \object{J0241-0326}, our results suggest periodic, brief increases of luminosity with a weighted mean period of 307$\pm$21~days, at a confidence level of 40\%, and peak-to-peak variations of 44$\pm$10~mmag. However, because of the low number of observations for this target, this should be confirmed with an improved dataset. Interestingly though, our absolute $K_s$-band photometry for this object is 0.1~mag fainter than its 2MASS magnitude, suggesting variability at even longer scales than studied in this work. The object \object{G196-3B}, on the other hand, displays a fairly sinusoidal phased curve corresponding to a weighted mean period of 442$\pm$7~days at a confidence level of 95\%, and peak-to-peak variations of $\pm$10~mmag. This period value is nearly a factor of two the astrometric period found by \citet{zo2014}. However, we are cautious about this result, because a similar period is also suggested by the periodograms for the comparison stars, albeit with p-values a factor of 100 higher. It is also unclear which physical effect could produce the observed variation (if confirmed) at such a long scale, although the fact that the found possible period is nearly twice the astrometric period reported by \citet{zo2014} suggests that the astrometric period could be driven by the photometric variability. Hence, for both \object{J0241-0326} and \object{G196-3B}, follow-up observations are required to confirm the possible variability suggested by the present work, and to study its physical origin.

\begin{acknowledgements}
     
We thank J.~A. Caballero and P. Miles-P\'aez for useful discussions, and R. Hueso for providing information on the photometric variability of the giant planets in our Solar System. This research was funded by the Spanish \emph{Plan Nacional de Astronom\'{\i}a y Astrof\'{\i}sica} through project AYA 2011-30147-C03-03.\\

This publication makes use of data products from the Two Micron All Sky Survey (2MASS), which is a joint project of the University of Massachusetts and the Infrared Processing and Analysis Center/California Institute of Technology, funded by the US National Aeronautics and Space Administration and National Science Foundation.\\

This work benefitted from the use of the SIMBAD database and the Vizier Catalogue Service, both operated at the CDS (Strasbourg, France); and of NASA's Astrophysics Data System (ADS). We used the following Virtual Observatory tools: STILTS \citep{taylor2006} and TOPCAT \citep{taylor2005}, both developed by Mark Taylor at Bristol University; and Aladin, developed at the CDS. We also used Starlink's PERIOD package, as well as the periodogram services from the NASA Exoplanet Archive, operated by the California Institute of Technology.

\end{acknowledgements}


\bibliographystyle{aa} 
\bibliography{references}


\end{document}